\documentclass[prx,floatfix,showpacs,superscriptaddress,longbibliography,twocolumn]{revtex4-1}

\usepackage[utf8x]{inputenc}

\usepackage[USenglish]{babel}

\usepackage{amsmath,amssymb,amsfonts}

\usepackage{bm}

\usepackage{epsfig}
\usepackage{subfigure}

\usepackage[colorlinks=true,linkcolor=blue,citecolor=red]{hyperref}

\newcommand{\equ}[1]
{Eq.~(\ref{#1})}

\newcommand{\figu}[1]
{Fig.~\ref{#1}}

\newcommand{\secu}[1]
{Sec.~\ref{#1}}

\newcommand{\appu}[1]
{App.~\ref{#1}}

\def\=={\equiv}

\def\cG0{{\cal G}_0} 
\def\cG{{\cal G}}

\def\=={\equiv}

\newcommand{\be}{\begin{equation}}
\newcommand{\ee}{\end{equation}}

\begin{document}

\date{\today}

\author{Francesco~Grandi}
\affiliation{Department of Physics, University of Erlangen-N\"urnberg,
  91058 Erlangen, Germany}

\author{Martin~Eckstein}
\affiliation{Department of Physics, University of Erlangen-N\"urnberg,
  91058 Erlangen, Germany}

\title{Fluctuation control of non-thermal orbital order}

\begin{abstract}
Orbitally ordered states exhibit unique features which make them a promising platform for exploring the ultrafast dynamics of long-range order in solids: Their free energy typically has multiple discrete minima, and electric laser fields or selectively excited phonons can exert effective forces that may be used to steer the order parameter through these free energy landscapes. Moreover, their free energy strongly depends on fluctuations, and in some cases restoring forces close to a minimum are exclusively of entropic origin (order-by-disorder mechanisms). This can open pathways to control the dynamics of the order parameter via non-thermal fluctuations. In this work, we study the laser-induced non-equilibrium dynamics in a $120^\circ$ compass model, using time-dependent Ginzburg-Landau theory. We analyze protocols to switch the order parameter between equivalent configurations, with a focus on the interplay between the external force due to the driving field, and the non-thermal entropic forces. In particular, we find that remanent non-thermal fluctuations after some excitation can stabilize the high-symmetry phase even when the homogeneous potential has retrieved its low-temperature form, which facilitates laser-induced switching. 
\end{abstract}
\maketitle


\section{Introduction} \label{secI}

The manipulation of matter with light \cite{Basov2017,review_giannetti}  gives new perspectives on both fundamental theoretical questions and future technological applications, such as magnetic memory devices that operate at picosecond speed \cite{RevModPhys.82.2731}. Particularly appealing is the ultrafast manipulation of multi-minima free-energy landscapes to control competing phases or to reach hidden states along non-adiabatic pathways \cite{Fausti2011a, PhysRevX.10.021028, Stojchevska177, Ichikawa2011, Forst2011PRB,Li2018}. Such non-adiabatic transitions can be driven by generalized forces on various order parameters, as obtained by the nonlinear effect of the electric laser-field or selectively excited phonons on lattice distortions \cite{Forst2011,PhysRevB.89.220301}, superconducting pairing \cite{Murakami2017, Babadi2017, Kennes2017_NatPhys, PhysRevX.10.031028}, artificial magnetic fields \cite{Nova2017}, the Coulomb interaction \cite{PhysRevLett.115.187401,Kaiser2014_SciRep,PhysRevB.103.L041110}, magnetic exchange interactions \cite{PhysRevLett.103.097402,Mentink2015}, and more. A more indirect pathway to control the dynamics of an order parameter relies on the excitation of fast degrees of freedom, such as electronic variables, which then modify the phenomenological free energy density $f(\phi_0,T,...)$ of the order parameter $\phi_0$. Free energies which depend on the fast degrees of freedom via few coarse-grained variables only, such as an effective electronic temperature $T$ or a photo-doping density, can indeed explain many observations regarding light-induced phase transitions \cite{Yusupov2010_NatPhys,Schaefer2010, Huber2014,Beaud2014}.

On the other hand, assuming that $f(\phi_0,T)$ is  determined by a single effective temperature even out of equilibrium would require all degrees of freedom apart from $\phi_0$ to be in a quasi-equilibrium state, which is clearly approximate. For example, recent observations show that the charge-density wave (CDW) order in photo-excited  TbTe$_{3}$ apparently evolves in the low-temperature potential, even if the electrons are still found to be above the equilibrium $T_c$ \cite{maklar2020nonequilibrium}. In general, degrees of freedom which are not equilibrated should be included explicitly in a phenomenological description \cite{Zong2019_NatPhys,PhysRevLett.123.097601,Kogar2020_NatPhys}. These non-thermal fluctuations can renormalize the potential for $\phi_0$ in a different way as compared to a temperature change. In particular, this includes the long-wavelength spatial fluctuations of the order parameter \cite{PhysRevB.101.174306}, which, e.g.,  are found to delay the recovery of the symmetry-broken state after an excitation for an $O(N)$-symmetric order parameter. 

A particularly important effect of fluctuations on the non-equilibrium dynamics can be anticipated when fluctuations already have a qualitative effect on the free energy landscape in equilibrium. This is the case for systems which follow the so-called order-by-disorder mechanism. In this scenario, the classical ground state potential has a continuous symmetry, which is reduced to the lattice point group by fluctuations, i.e., by purely entropic forces. The concept was  introduced initially to describe the order induced by quantum and thermal fluctuations in otherwise frustrated spin models \cite{PhysRevLett.62.2056}, and it is realized in particular in a class of spatially anisotropic spin models known as compass models \cite{RevModPhys.87.1}. Such models are relevant for the description of orbital degrees of freedom in transition metal compounds \cite{kugel_khomskii_1973} and for applications in orbitronics \cite{PhysRevLett.95.066601} or topological quantum computing \cite{KITAEV20032}. The time-dependent control of the orbital state in solids might be achieved via an indirect coupling of the orbitals with the electromagnetic field \cite{eckstein2017designing}, or via their coupling to the lattice \cite{Polli2007,Miller2015,PhysRevB.95.195405,PhysRevLett.125.216404}. A measurement of the momentum-dependent order parameter fluctuations could then be done using time-resolved inelastic X-ray scattering (trRIXS) \cite{PhysRevB.99.104306, Mitranoeaax3346, Mitrano2020, PhysRevResearch.2.023110}.

The  focus of the present study is to investigate the effect of fluctuations on the non-equilibrium order parameter dynamics in a paradigmatic model for orbitally-ordered systems. We consider the $120^{\circ}$ compass model, an orbital-only model representing Mott insulators with one electron (or one hole) in an $\text{e}_{\text{g}}$ orbital doublet. In spite of the strong directionality of the $\text{e}_{\text{g}}$ orbitals, the classical order parameter potential at zero temperature has a continuous symmetry, which is reduced to the cubic point group symmetry via the order-by-disorder mechanism. We investigate the role of fluctuations on the ultrafast quench and recovery after various excitation protocols. For a simple temperature quench of the fast degrees of freedom, non-thermal fluctuations tend to stabilize the high symmetry phase even when the temperature of the fast degrees of freedom goes back to the original value, similar as for the $O(N)$-symmetric order parameter \cite{PhysRevB.101.174306}. As a second excitation protocol we consider a time-dependent electric field, which can provide an effective coherent force on the order parameter, and thus allows the system to switch from one orbital state to another. We show how the spectrum of the fluctuations evolves along the switching trajectory, emphasizing the interplay between the external force  due to the driving field, and the entropic forces. Finally, combining the coherent driving with a temperature quench permits to switch the orbital state much faster than at fixed temperature, similar to heat-assisted switching in magnetic memory devices. In this protocol, the transient trapping of the system in the high symmetry state due to non-thermal fluctuations is essential to facilitate an efficient switching process.

The manuscript is structured as follows: In \secu{secII}, we introduce the model and a time-dependent Ginzburg-Landau formalism implemented to take into account the fluctuations of the order parameters. \secu{secIII} presents the main results of this work, particularly the effect of a temperature quench on the dynamics of the order parameter (\secu{secIIIquench}), a coherent control of the order parameter through the manipulation of the orbital exchange coupling (\secu{secIIIswitch}) and the combination of the two time-dependent perturbations (\secu{secIIIquenchswitch}). In \secu{secIV}, we discuss possible experimental probes and establish a connection between the correlation functions obtained within our formalism and the structure factors measurable with trRIXS. Finally, \secu{secV} concludes the work.

\section{Model and formalism} \label{secII}

\subsection{Ginzburg-Landau formalism for the $120^\circ$ compass model } \label{secIIa}

In the following, we will analyze the $120^{\circ}$ compass model on a cubic lattice as a specific representative for the general class of orbital-only compass models \cite{RevModPhys.87.1}. This model is obtained from an underlying electronic Hamiltonian for a single electron or hole in the $\text{e}_{\text{g}}$ orbital doublet, after freezing both charge degrees of freedom (which is justified in the Mott regime) and the electron spin. Freezing the spin can be achieved by strong magnetic fields, or alternatively, the dynamics of the spin may be neglected to first order at temperatures where magnetic order is absent while orbital order remains intact \cite{Nussinov_2004}. 

We define the orbital model in terms of a $\tfrac12$-pseudospin $\vec{\tau}=(\tau^1,\tau^2,\tau^3)$ acting in the  orbital e$_{\text{g}}$-doublet. Using a basis where the eigenstates of the $\tau^3$ operators are the $|z^2-y^2\rangle$ and $|3x^2-r^2\rangle$ orbitals, $P^{x}=\frac{1}{2}(1-\tau^3)$ is the projector on the $|3x^2-r^2\rangle$ orbital. The projectors $P^{\gamma}$ on the orbital $|3\gamma^2-r^2\rangle$ for $\gamma=x,y,z$ are then rotated with respect to each other by $120^\circ$ around the $\tau^2$-direction  in pseudospin space, i.e., $P^{\gamma}=\tfrac12( 1- n^\gamma_i\tau^i)$, where summation over double indices is implied, and $n^\gamma_i$ are the components of the unit vectors $\hat n^x=(0,0,1)$, $\hat n^y=(\sqrt{3},0,-1)/2$, $\hat n^z=(-\sqrt{3},0,-1)/2$. The orbital exchange Hamiltonian for a bond $(\mathbf{R},\mathbf{R}+\mathbf{e}_{\gamma})$ along a given direction $\gamma=x,y,z$ is $-\bar J_\gamma P^{\gamma}_{\mathbf{R}}P^{\gamma}_{\mathbf{R}+\mathbf{e}_{\gamma}}$, favoring both electrons in the orbital $|3\gamma^2-r^2\rangle$ with the biggest overlap along the bond. The total Hamiltonian is therefore written as
\be \label{ham}
H = - \sum_{\mathbf{R}} \Big[
\frac{1}{4}\sum_{\gamma} \bar{J}_{\gamma}
n^{\gamma}_{i}n^{\gamma}_{j} \tau_{\mathbf{R}}^i \tau_{\mathbf{R}+\mathbf{e}_{\gamma}}^j-\sum_{\gamma} \bar{J}_{\gamma}n^{\gamma}_{i} \tau_{\mathbf{R}}^i
\Big],
\ee
up to irrelevant constants. In equilibrium, the exchange constants $\bar J_{\gamma}$ for $\gamma=x,y,z$ are equal, but they can become different when external fields are applied (see below). In particular, this implies that the second term in the Hamiltonian, which corresponds to a pseudo-spin magnetic field, is absent in equilibrium, because $\sum_\gamma n_{\gamma}=0$. 

The classical ground state of model \equ{ham} is ferro-orbitally ordered, and fully degenerate as function of the angle of the ordered moment in the $\left( \tau^{3}, \tau^{1} \right)$ plane. However, as soon as thermal or quantum fluctuations are taken into account, the degeneracy is broken, leading to the stabilization of six degenerate minima which correspond to equivalent ferro-orbital order polarizations that are mapped onto each other by the cubic point group operations \cite{PhysRevB.59.6795,Brink_2004,Biskup2005}. This order-by-disorder mechanism can be understood already at the level of spin-wave theory, because the excitation spectrum, and thus the entropic contribution to the free energy at nonzero temperature, depends on the direction of the ordered orbital moment. 

To describe the physics of model \eqref{ham} within the time-dependent Ginzburg-Landau formalism, we write a free-energy functional which is consistent with the continuum limit of the Hamiltonian \eqref{ham}, 
\begin{align}
\label{free_en}
\mathcal{F} [ \phi] 
&= 
\int d^{3} \mathbf{x}
\Big[\, \frac{r(T)}{2}\sum_{\gamma} J_\gamma  n^\gamma_{i}  n^\gamma_{j} \phi_j \phi_i
-
\sum_{\gamma} J_\gamma  n^\gamma_{i}  \phi_i
\nonumber\\
&
+
u (\phi_i\phi_i)^2
+\frac{K}{2} \sum_{\gamma}
J_\gamma
n^\gamma_{i}  n^\gamma_{j}  (\partial_\gamma \phi_j) (\partial_\gamma  \phi_i)
\Big].
\end{align}
Here  $(\phi_{1} \left( \mathbf{x}, t \right), \phi_{2} \left( \mathbf{x}, t \right))$ is a two-dimensional order parameter that corresponds to the coarse-grained pseudospin components ($\tau^{3},\tau^{1})$. From now on, we restrict the orbital indexes to the values $1,2$, imply summation over double orbital indices, and redefine the unitary vectors $\hat n^x=(1,0)$, $\hat n^y=(-1,\sqrt{3})/2$ and $\hat n^z=(-1,-\sqrt{3})/2$ by omitting the irrelevant $\tau^2$ component. The  coefficient of the quadratic contribution in Eq.~\eqref{free_en} is proportional to a temperature-dependent factor $r(T)= - 2 (1 - \frac{T}{T_{c}})$ which changes sign at the critical temperature $T_{c}$. The quartic term $u$ implements a soft-spin approximation and prevents the order parameter to grow arbitrarily large. The term proportional to $K$, which has the form of a gradient approximation to the exchange interaction in Eq.~\eqref{ham}, is the free energy contribution from spatial order parameter fluctuations. The couplings $J_{\gamma}$ in the free energy are related to $\bar{J}_{\gamma}$ in \equ{ham} through some coarse-graining, and must thus be understood as phenomenological parameters. However, one can assume that they can be modified by external fields in the same way as the bare parameters $\bar{J}_{\gamma}$. In particular, $J_{x} = J_{y} = J_{z} \equiv J$ in equilibrium, which implies that the free energy density arising from the  first three terms in Eq.~\eqref{free_en} becomes $\frac{3}{4} r(T)J \phi_i\phi_i+ u(\phi_i\phi_i)^2$, corresponding to the standard $O(2)$-symmetric $\phi^4$ theory \cite{goldston_gauss}. In contrast, the term proportional to $K$ is only invariant under the cubic point group, implementing again the order-by-disorder scenario.

\subsection{Time-evolution using model A dynamics} \label{secIIb}

We compute the time evolution assuming an over-damped dissipative dynamics (model A dynamics), which is applicable for anisotropic magnets and systems without conserved fields \cite{RevModPhys.49.435,PhysRevB.103.035116}. The dissipative equations of motion  read
\be \label{langevin}
\dot{\phi}_{i} \left( \mathbf{x}, t \right)
= - \Gamma \frac{\delta \mathcal{F}}{\delta \phi_{i} \left( \mathbf{x}, t \right)} + \eta_{i} \left( \mathbf{x}, t \right) \;,
\ee
where $\Gamma> 0$ is the relaxation rate, and $\eta_{i} \left( \mathbf{x}, t \right)$ is a Gaussian white noise term that takes into account the effect of the fast degrees of freedom, characterized by the moments
\be \label{corr}
\begin{split}
& \langle \eta_{i} \left( \mathbf{x}, t \right) \rangle = 0 \;, \\
& \langle \eta_{i} \left( \mathbf{x}, t \right) \eta_{j} \left( \mathbf{x'}, t' \right) \rangle = 2 T \Gamma \delta_{i,j} \delta \left( \mathbf{x} - \mathbf{x'} \right) \delta \left( t - t' \right) \;.
\end{split}
\ee
The parameter $T$ in \equ{corr} is the temperature of the fast degrees of freedom, which are integrated out in the coarse-graining under the assumption that they instantly thermalize on the  timescale of the order parameter dynamics. The noise temperature $T$ will therefore be assumed to be equal to the temperature in the coefficient $r(T)$, even in an excitation protocol where $T$ becomes time-dependent (see Sec.~\ref{secIII}). 

With the Fourier transformation of the order parameter we define the correlation functions $D_{i j} \left( \mathbf{k}, t \right) = \langle \phi_{i} \left( \mathbf{k}, t \right) \phi_{j} \left( -\mathbf{k}, t \right) \rangle$ in reciprocal space, and the total number of excitations in a particular channel,
\be \label{exc_num}
n_{i j} \left( t \right) = \frac{1}{ 8\pi^{3}} \int_{-\Lambda}^{\Lambda} d k_{1} \int_{-\Lambda}^{\Lambda} d k_{2} \int_{-\Lambda}^{\Lambda} d k_{3} \ D_{i j} \left( \mathbf{k}, t \right).
\ee
Here $\Lambda$ is an ultraviolet cutoff. 
Note that the correlation matrix is symmetric, $n_{i j}=n_{j i}$. The dynamical model given by Eqs.~\eqref{langevin} and \eqref{corr} can then be recast as a Fokker-Planck-Smoluchowski equation for the probability distribution of the fields $\phi_{i} \left( \mathbf{x}, t \right)$. Under the assumption that the probability distribution is Gaussian, we obtain a closed set of equations for the correlation function and the statistically averaged fields $\bar{\phi}_{i} \left( t \right) = \langle \phi_{i} \left( \mathbf{x}, t \right) \rangle$, 
\begin{align} 
\label{dyn_eq}
\dot{\bar{\phi}}_{i} (t)
&= - \Gamma \Big[A^{\text{eff}}_{i j} \bar{\phi}_{j} 
-\sum_\gamma J_\gamma n^\gamma_i 
\Big],
\\
\label{dyn_eq_corr}
\dot D_{ij}(\mathbf{k}, t)
&= 
\Gamma \Big[
2\delta_{ij} T -
M_{il}D_{lj}(\mathbf{k}, t) - D_{il}(\mathbf{k}, t)M_{lj}
\Big],
\end{align}
where 
\begin{align}
\label{r_param_eff}
A^{\text{eff}}_{ij}&= 
A^{0}_{ij}
+
4u\big[\delta_{ij}
\sum_l(\bar{\phi}_{l}^{2} + n_{ll})
+
2 n_{ij}\big],
\\
\label{m_matrix}
M_{ij}&= A^{\text{eff}}_{i j}+\kappa_{i j} +8u \bar{\phi}_{i} \bar{\phi}_{j} \;,
\end{align}

with the bare coupling matrix
\be \label{r_param_bare}
A^{0}_{ij}(T) = r(T)\sum_\gamma J_\gamma n^\gamma_i  n^\gamma_j \;,
\ee
and the distortion contribution
\be \label{distort_coeff}
\kappa_{ij} = K\sum_\gamma J_\gamma n^\gamma_i  n^\gamma_j k_\gamma^2 \;.
\ee
The second term in the first equation acts like a  bare external force on the order parameter,  
\begin{align}
\label{fj0} 
f^0_i=\sum_\gamma J_\gamma n^\gamma_i.
\end{align}
The equations are solved numerically on an equidistant grid of k-points; $N_{k} \sim \left( 300 \right)^{3}$ to $\left( 400 \right)^{3}$ is sufficient to obtain converged results.

To analyze the results below, we will parametrrize the order parameter ($\bar{\phi}_{1},\bar{\phi}_{2})$ in polar coordinates, with amplitude and phase defined as
\be
\begin{pmatrix}
\bar \phi_1
\\
\bar \phi_2
\end{pmatrix}
=
R
\begin{pmatrix}
\sin(\varphi)
\\
-\cos(\varphi)
\end{pmatrix} \;.
\ee
The zero-temperature stationary solution of Eqs.~\eqref{dyn_eq} and \eqref{dyn_eq_corr} implies $n_{ij}=0$, and reflects the $\text{O} ( 2 )$ symmetry of the problem. For nonzero temperatures below the critical value, there are six equivalent stationary solutions characterized by the same amplitude $R$ but different phase $\varphi = \pi/6, \pi/2, 5\pi/6, 7\pi/6, 3\pi/2$ and $11\pi/6$, each separated by an angle $\pi/3$. More details about the equilibrium solution can be found in \appu{app_a}.

In the following, we take the parameters $J_{x} = J_{y}= J_{z} = 5$,  $u=1$, $K=1$, $T_{c} = 0.5$, $\Lambda = \pi$ and $\Gamma = 0.5$. At zero temperature, the free energy landscape (as well as $\Gamma$ and $\Lambda$) is the same as for the totally $O(N)$-symmetric model defined in Ref.~\cite{PhysRevB.101.174306} when $N=2$. Unless otherwise specified, the equilibrium temperature of the system is set to $T = 0.125$ and, without any loss of generality, we initially place the state of the system at  $\varphi = \pi/6$. 

\subsection{Effective potential for the homogeneous order parameter} \label{secIIc}

Looking at Eq.~\eqref{dyn_eq}, one can formally write the dynamics for the homogeneous order parameter $\bar \phi$ in terms of an effective potential $\bar F(\bar \phi)$, such that
\begin{align}
\dot{\bar\phi}_i &= -\Gamma \frac{\partial \bar F(\bar \phi)}{\partial \bar\phi_i}.
\end{align}
The corresponding free energy  $\bar F$,
\begin{align}
\label{fbar}
\bar F(\bar \phi)
&=
\frac{1}{2}
A^0_{ij}(T)\bar \phi_i\bar \phi_j
+
u (\bar \phi_i\bar \phi_i)^2
-
f_i^0 \bar \phi_i
+
F_{\text{fl}}(\bar \phi),
\end{align}
is given by the homogenous part of Eq.~\eqref{free_en} and a quadratic fluctuation potential $F_{\text{fl}}(\bar \phi)=\tfrac12 A_{ij}^{\text{fl}}\bar \phi_i \bar \phi_j$, with the fluctuation matrix
\begin{align}
\label{Afl}
A_{ij}^{\text{fl}}
=
4u\big[\delta_{ij}\sum_l n_{ll} +  2n_{ij}\big].
\end{align}
During the dynamics, the fluctuations change in a nontrivial way, and the  matrix $A^{\text{fl}}$ will provide a convenient way to quantify their effect on the dynamics of the order parameter.

\section{Results} \label{secIII}

\begin{figure}
\centering \includegraphics[width=0.5\textwidth]{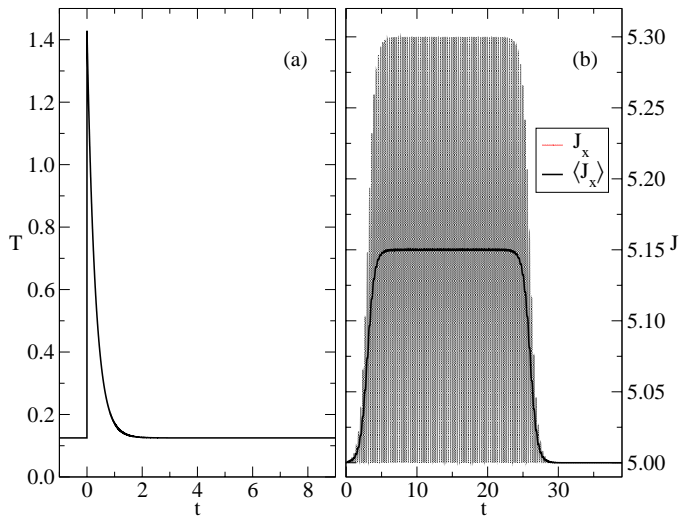}
\caption{ (a) Time-dependent profile of the temperature quench \eqref{temp_quench}, for $\tau_{QP} = 0.3$, $T_{\text{s}} = 0.125$ and $T_{\text{f}} = 1.428$. (b) Time dependence of the exchange parameter $J_x$ used in the switching protocol [Eq.~\eqref{jprofile} with $\delta J=0.3$, $J_x^0=5$, $\bar{\Omega} = 10$ and $t_{p} \sim 30$] together with its moving average $\langle J_{x} \rangle$.}
\label{fig_quench_J_vs_time}
\end{figure}

In the following, we analyze three different time-dependent excitation protocols: (i) A short temperature quench with subsequent recovery, implemented by a time-dependent temperature $T(t)$ in Eqs.~\eqref{corr}, \eqref{dyn_eq_corr}, and \eqref{r_param_bare} (see \figu{fig_quench_J_vs_time}(a)), (ii), a time-dependent modulation of the the renormalized exchange interaction $J_{x} \left( t \right)$, as shown in \figu{fig_quench_J_vs_time}(b), and (iii), a combination of (i) and (ii). The excitation protocol (i) is used frequently in the literature to describe the laser-induced heating of the fast degrees of freedom, such as electrons, which couple directly to the light pulse and are assumed to thermalize instantly on the timescale of the order-parameter dynamics. The exponential recovery of the initial temperature after the pulse reflects the energy dissipation to degrees of freedom which do not strongly affect the order parameter (such as acoustic phonons). As mentioned,  $T(t)$ enters both the free energy potential \eqref{free_en} and the noise \eqref{corr}.

The excitation protocol (ii) finds its rationale in the possibility of controlling the time-dependent profiles of $\bar{J}_{x} \left( t \right), \bar{J}_{y} \left( t \right)$ and $\bar{J}_{z} \left( t \right)$ separately with suitable electric fields directed along given lattice directions. Orbital exchange interactions can, e.g., be changed directly by manipulating the exchange path with the electric field of a laser. Explicit expressions for the exchange couplings in the field-driven two-orbital Hubbard model have been obtained from a time-dependent Schrieffer-Wolff transformation \cite{eckstein2017designing}, indicating that few-percent changes of $J$ may be possible in experiment. One can also envision other symmetry-allowed processes, such as driving infrared-active phonons which are nonlinearly coupled to the orbital pseudospin. In this study, we assume that the effective parameters $J_{x} \left( t \right), J_{y} \left( t \right)$ and $J_{z} \left( t \right)$ in the coarse-grained free energy are controlled in the same way as the microscopic parameters. This enables a coherent control of the order parameter,  with a switching between different free energy minima.

Finally, protocol (iii) is motivated by the aim to explore how the fluctuation-induced change of the free energy cooperates or competes with the explicit control by the modulation of the exchange couplings. In practice, any implementation of the coherent control protocol (ii) will go together with a transient heating of the fast degrees of freedom.

\subsection{Temperature quench} \label{secIIIquench}

\begin{figure}
\centering \includegraphics[width=0.5\textwidth]{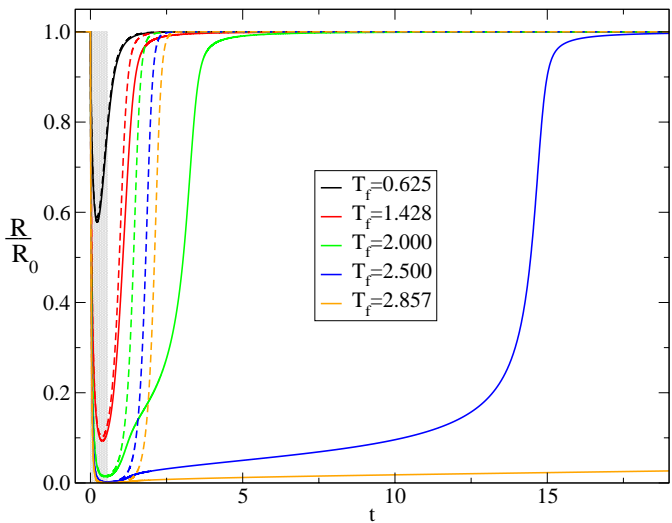}
\caption{Solid lines show the time-dependent change of the amplitude of the order parameter (normalized by the equilibrium value $R_{0}$) during  temperature quenches to different maximum temperatures $T_{\text{f}}$. For all simulations, the initial temperature is $T_{\text{s}} = 0.125$, and $\tau_{QP}=0.3$. Dashed lines correspond to the dynamics of the order parameter computed with the free energy \eqref{fbar-T} without including the fluctuations (see text). The shaded region indicates times for which $T>T_c$ for the quench at $T_{\text{f}} = 2.500$.}
\label{fig_temp_quench_rT_comp}
\end{figure}

In this section, we focus on the effects of a temperature quench on the dynamics of the system. We assume the time-dependent temperature
\be \label{temp_quench}
T \left( t \right) = T_{\text{s}} + \left( T_{\text{f}} - T_{\text{s}} \right) \theta \left( t \right) e^{-t/\tau_{QP}} \;,
\ee
where $\theta \left( t \right)$ is the Heaviside step function, $T_{\text{s}}$ the pre-pulse temperature, $T_{\text{f}}$ the maximum value of the temperature reached at $t=0^{+}$, and $\tau_{QP} = 0.3$ the relaxation rate to $T_{\text{s}}$ (see \figu{fig_quench_J_vs_time}(a)). The dynamics following such quenches is shown in \figu{fig_temp_quench_rT_comp}. The amplitude $R(t)$ of the order parameter is suppressed by the temperature increase and subsequently recovers to the initial value. Upon increasing $T_{\text{f}}$, the order parameter is suppressed to a value closer to zero, and the recovery time increases. 

\begin{figure}
\centering \includegraphics[width=0.5\textwidth]{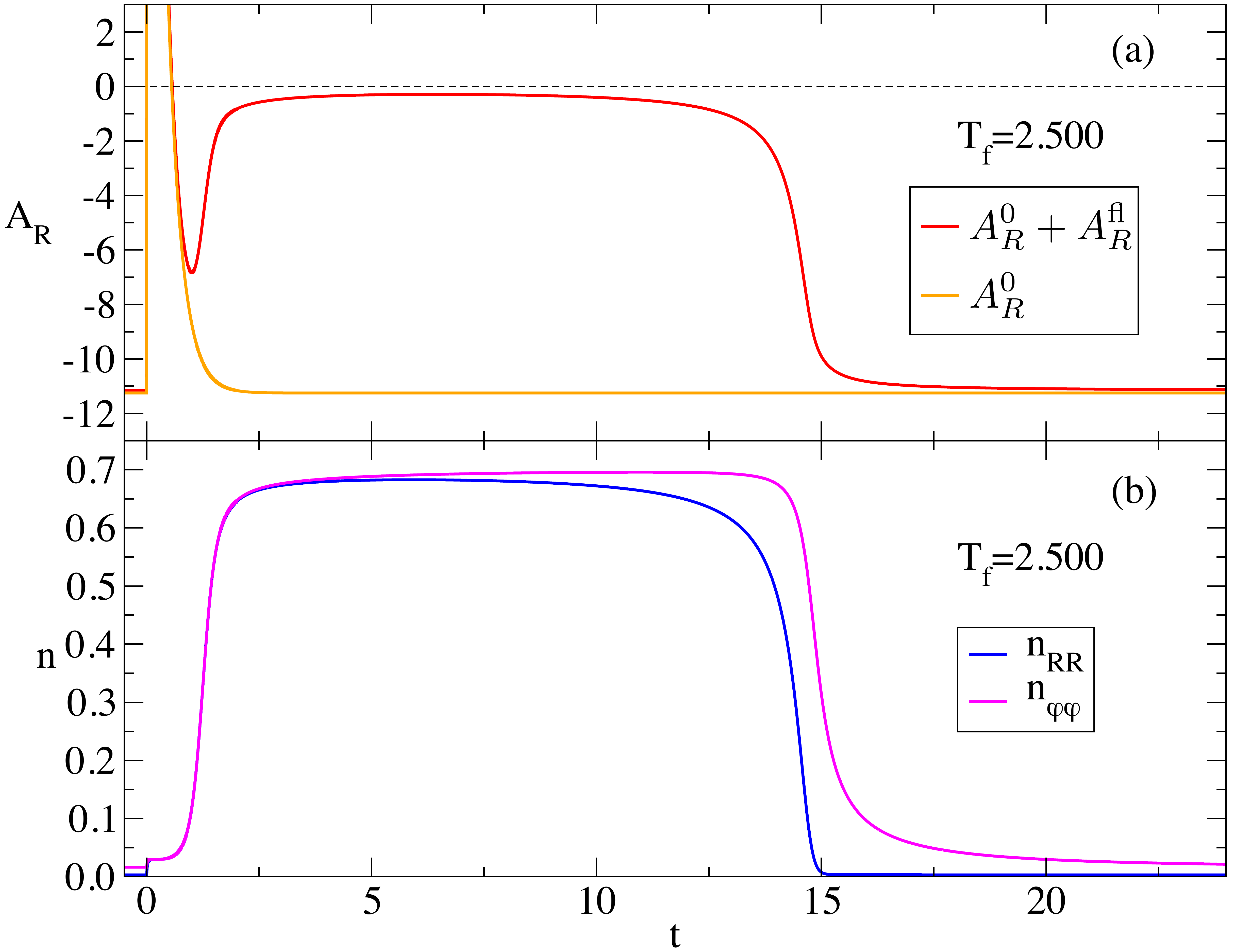}
\caption{ (a) Analysis of the free energy \eqref{fbar-T} close to $R=0$: The bare coefficient $A^0_R$ of the quadratic term and the fluctuation-renormalized coefficient $A^0_R+A^{\text{fl}}_R$ are plotted as function of time for the quench with $T_{\text{f}}=2.500$ shown in \figu{fig_temp_quench_rT_comp}. The dashed line indicates the zero of the $A_{R}$ parameter. (b) Excitation density in polar coordinates  for the same quench. The mixed component $n_{\varphi R}$ remains equal to zero during the whole dynamics.}
\label{fig_n_polar_vs_time_bf_05}
\end{figure}

The phase $\varphi = \pi/6$ remains constant after the temperature quench.  To characterize the dynamics, we  can therefore analyze the effective potential \eqref{fbar} at fixed angle $\varphi$ as a function of $R$. With $J_x=J_y=J_z\equiv J$,
\begin{align}
\label{fbar-T}
\bar F(R)
&=
\tfrac12
[A^0_R(t)+A^{\text{fl}}_R(t)] R^2
+
u R^4,
\end{align}
where $A^0_R(t)= \frac{3}{2} r(T(t))J$, and the fluctuation component $A^{\text{fl}}_R(t)$ is given by the projection of the matrix $A_{ij}^{\text{fl}}$ in Eq.~\eqref{Afl} along the unit vector $\hat e^R = (\sin(\pi/6),-\cos(\pi/6))$ in radial direction in orbital space. If one would consider only the dynamics generated by the bare potential ($\dot R = -\Gamma \partial_R \bar F(R)$, with $A^{\text{fl}}=0$), one  obtains a somewhat similar result as for the full solution (see dashed lines in Fig.~\ref{fig_temp_quench_rT_comp}). In this case, the delayed recovery after strong quenches has a trivial interpretation: The quench brings the order parameter to a small value close to $R=0$. After some time $t_{\text{min}}$ when the initial temperature is recovered (which is very early on the scale of the plot, see shaded region in Fig.~\ref{fig_temp_quench_rT_comp}), the order parameter undergoes an exponential growth phase $R(t)\sim R_{\text{min}}e^{A^0 (t-t_{\text{min}})}$ dominated by the quadratic term in $\bar F$, starting from the small value $R_{\text{min}}$ at time $t=t_{\text{min}}$. The stronger the quench, the smaller $R_{\text{min}}$, and the longer it takes for the order parameter to leave the unstable region. However, it is clear from the comparison of the dashed and full lines in Fig.~\ref{fig_temp_quench_rT_comp}, that this argument would largely underestimate the delay in the recovery of the order parameter.  The latter is therefore described only if the fluctuations are taken into account. This delayed recovery is qualitatively the same as given for the $O(N)$-symmetric model \cite{PhysRevB.101.174306}. To further analyze the behavior, we plot in \figu{fig_n_polar_vs_time_bf_05}(a) both the bare component $A^0_R$ and the fluctuation-renormalized coefficient $A^0_R+A_R^{\text{fl}}$ [c.f.~Eq.~\eqref{fbar-T}] of the quadratic term in $\bar F(R)$. The comparison shows that while the bare potential would be already unstable for most of the time ($A^0_R<0$), the fluctuations tend to stabilize the high-symmetry state. 

Finally, it is interesting to analyze the fluctuations themselves. For this purpose, we chose a representation using polar coordinates in orbital space,
\be \label{fluct_polar}
\begin{split}
& n_{\varphi \varphi} = \sin^{2} \left( \varphi \right) n_{22}+ \cos^{2} \left( \varphi \right) n_{11} + \sin \left( 2 \varphi \right) n_{12} \;, \\
& n_{R R} =  \sin^{2} \left( \varphi \right) n_{11}+ \cos^{2} \left( \varphi \right) n_{22} - \sin \left( 2 \varphi \right) n_{12} \;, \\
& n_{\varphi R} = \sin \left( \varphi \right) \cos \left( \varphi \right) \left( n_{11} - n_{22} \right) -\cos \left( 2 \varphi \right) n_{12} \;,
\end{split}
\ee
with $\varphi = \pi/6$. The components are obtained by projecting the fluctuation matrix $n_{ij}$ along the given directions in orbital space, e.g.,  $n_{RR}=n_{ij}e^R_i e^R_j $, with the unit vector $e^R =(\sin(\varphi),-\cos(\varphi))$ along the radial direction. In \figu{fig_n_polar_vs_time_bf_05}(b), a separation of the time-scales in the recovery of the radial and angular fluctuations ($n_{RR}$ and $n_{\varphi \varphi}$, respectively) becomes apparent. This observation simply reflects the fact that the free energy landscape is more shallow in the angular than in radial direction. A similar effect has been observed for the $O(N)$-symmetric model \cite{PhysRevB.101.174306}. 
In the present case, the model does not have a continuous symmetry at nonzero temperature, and the angular fluctuations are therefore stronger confined. A measurement of the fluctuation dynamics in the two directions (see also Sec.~\ref{secIV}) could allow to map out this nontrivial potential landscape arising from the order-by-disorder mechanism.


\subsection{Time-dependent control of the exchange parameters} \label{secIIIswitch}

\begin{figure}
\centering \includegraphics[width=0.5\textwidth]{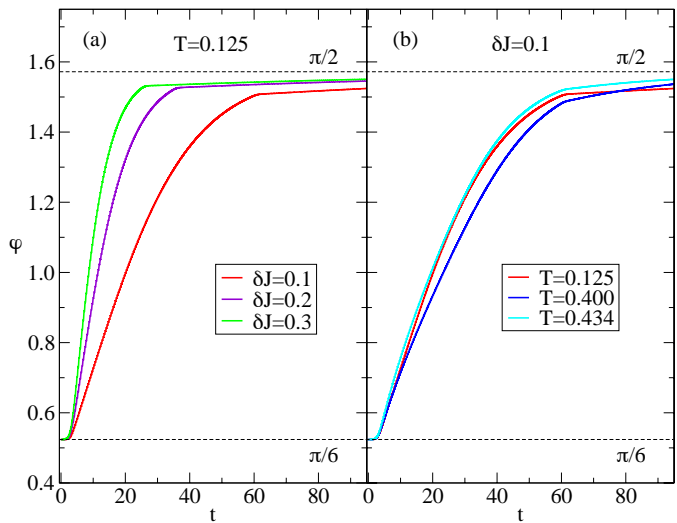}
\caption{ (a) Time-dependence of the angle $\varphi$ when the system evolves with the time-dependent exchange interactions shown in \figu{fig_quench_J_vs_time}(b), for several values of the variation $\delta J$. The pulse durations are $t_p \sim 65, 40, 30$ for $\delta J= 0.1, 0.2, 0.3$, respectively. (b) Same as panel (a), but for fixed pulse parameters $\delta J=0.1$ and $t_p \sim 65$, and different initial temperatures.
}
\label{fig_ang_n_cart_vs_time_comp}
\end{figure}

In this section, we describe the time evolution of the system during an orbital switching process, where the state moves from the free energy minimum at $\varphi = \pi/6$ to the one at $\varphi = \pi/2$. To achieve the switching, we modulate the value of the  exchange coupling $J_{x}$ as shown in \figu{fig_quench_J_vs_time}(b). The precise time profile is
\be
\label{jprofile}
J_x(t)=J_x^0 + \delta J \cos(\bar{\Omega} t)^2 s(t),
\ee
where $s(t)$ is a smooth envelope of duration $t_p$. The oscillatory time-dependence, with a large frequency $\bar\Omega$ is motivated by the shape one would obtain when the exchange couplings are controlled with oscillatory electric fields \cite{eckstein2017designing}. Clearly, the fast oscillations are almost irrelevant for the slow dynamics of the order parameter, and one could equally well replace $\cos(\bar{\Omega} t)^2 \to \tfrac12$.  The pulse is therefore determined by the equilibrium coupling $J_x^0$, the duration $t_p$, and the amplitude $\delta J$, which will generally taken to be of the order of few percent of $J$.

The first effect of this protocol is to make the pseudo-spin magnetic fields $f_i^0$ nonzero [c.f.~Eq.~\eqref{fj0}]. In addition, one has a direct modification of the quadratic terms $A_{ij}^0$ in the potential, and the indirect effect through the fluctuations. As shown in \figu{fig_ang_n_cart_vs_time_comp}(a), the given protocol can lead to a switch from one minimum to the other. As expected, with larger $\delta J$, the final minimum is reached faster. We now fix the pulse parameters $\delta J = 0.1$ and $t_p \sim 65$ and compare the dynamics for different temperatures (\figu{fig_ang_n_cart_vs_time_comp}(b)). The resulting switching time is non-monotonous as a function of $T$. As the temperature is increased, the switching first becomes slower (compare the curves for $T=0.125$ and $T=0.400$), and then again faster, as the  critical temperature is approached. As we will explain in the following, this non-monotonous behavior  again reflects the  order-by-disorder mechanism.

\begin{figure}[tbp]
\centering \includegraphics[width=0.45\textwidth]{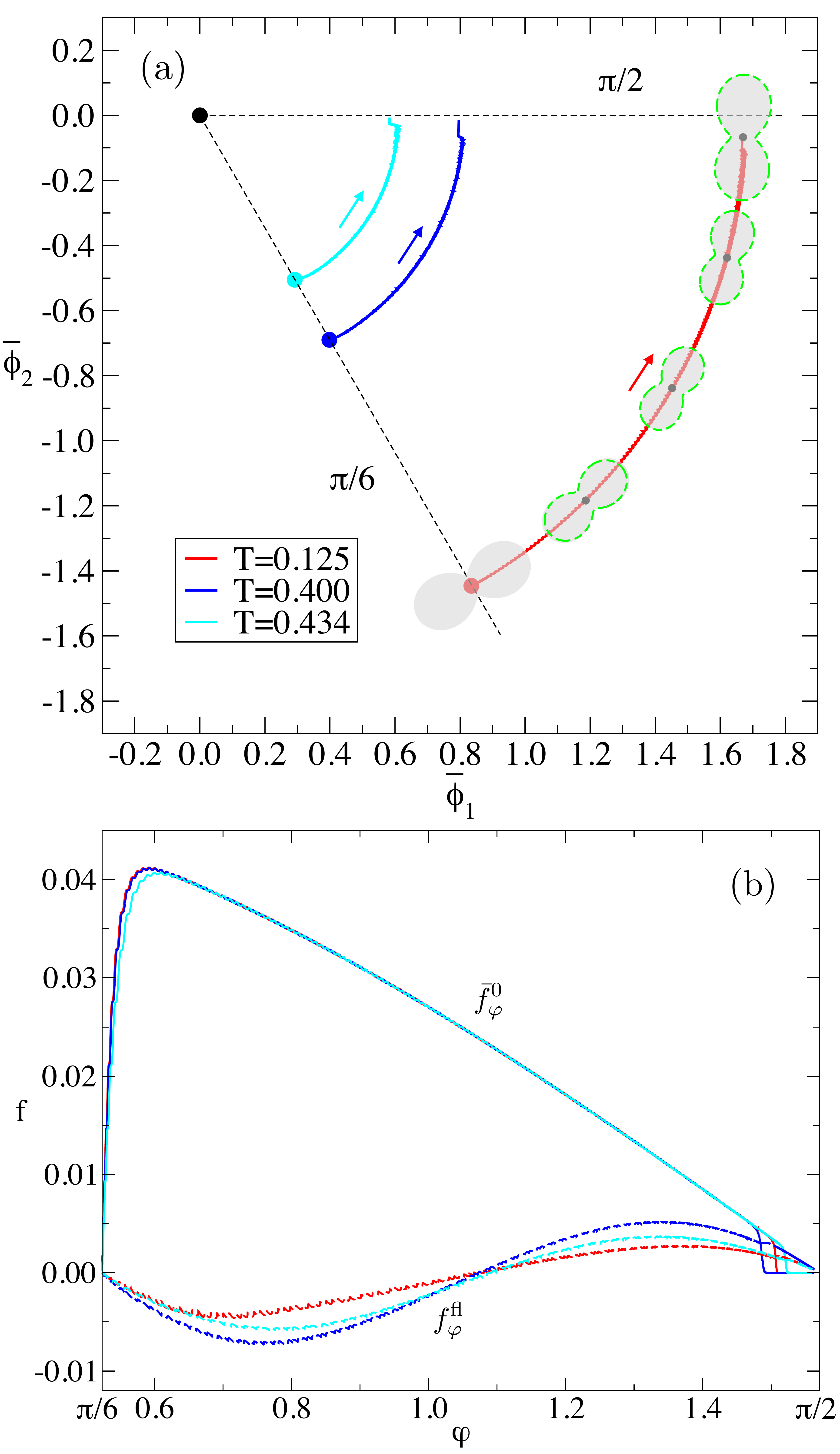}
\caption{ 
(a) Trajectories of the order parameter for the manipulation of $J_{x}$  at three different initial temperatures  [fixed pulse parameters $\delta J=0.1$ and $t_p \sim 65$ as in \figu{fig_ang_n_cart_vs_time_comp}(b)]. The arrows indicate the direction of motion. The shaded grey areas along the curve at $T=0.125$ show a polar representation $n(\varphi',t)$ of the non-equilibrium fluctuation matrix around different points $\varphi$ along the trajectory [see \equ{fluc_ang_distrib}]. The dashed green lines show the corresponding equilibrium fluctuations distribution for the same points in the order parameter plane. (b) Projections of the moving average of the external ($\bar{f}^{0}_{\varphi}$, solid lines) and fluctuating ($f^{\text{fl}}_{\varphi}$, dashed lines) contributions of the forces along the angular direction as function of the angle $\varphi$ for the three trajectories considered in panel (a).}
\label{fig_ang_trajec_diff_temp}
\end{figure}

The trajectories in the order parameter plane, corresponding to the switching protocols described in \figu{fig_ang_n_cart_vs_time_comp}(b), are shown in \figu{fig_ang_trajec_diff_temp}(a). The three curves, at different amplitude of the order parameter,   correspond to the three different temperatures. The non-monotonous behavior of the switching speed can now be understood by analyzing the different contributions to the effective force on the order parameter. For this purpose, we project the external force $f_i^0$ (\equ{fj0}) at each point $\varphi$ on the trajectory along the angular direction $e^\varphi=(\cos(\varphi),\sin(\varphi))$ in orbital pseudo-spin space,
\be \label{f0ang}
f^{0}_{\varphi} = f^{0} \cdot  e^\varphi = f_1^0 \cos(\varphi) + f_2^0 \sin(\varphi) \;,
\ee
and compare it with the force contribution from the fluctuating part of the potential \eqref{fbar},
\be
f_i^{\text{fl}}= - \partial F^{\text{fl}}/\partial \bar \phi_i = - A^{\text{fl}}_{ij}\bar\phi_j \;,
\ee
also projected along the angular direction $f^{\text{fl}}_{\varphi} = f^{\text{fl}} \cdot e^\varphi$. The result is shown in \figu{fig_ang_trajec_diff_temp}(b). Because $f^0$  depends on the rapidly oscillating  parameter $J_{x}$, see \equ{f0ang}, we show the moving average $\bar{f}^{0}_{\varphi}$ over an oscillation period. In this representation, a positive force contributes to the drag in the angular direction from $\pi/6$ to $\pi/2$, while a negative force opposes the motion.  The bare $\bar{f}^{0}_{\varphi}$, which directly depends on the exchange constants, is almost unchanged with temperature. For phases close to $\pi/6$, the fluctuating force is negative, indicating that the fluctuations would pull the system back to the original minimum, while for $\varphi$  larger than $\pi /3 \sim 1.0472$, the fluctuation force pulls the system towards the new minimum, and the system would evolve towards $\varphi=\pi/2$ even without the external force $f^{0}_{\varphi}$.  Apparently, it is this fluctuation force which reflects the non-monotonous dependence of the switching on $T$. This would be consistent with the order-by-disorder mechanism:  For small $T$, the increase of the temperature increases the fluctuation-induced forces, while closer to the phase transition, where the amplitude $R$ of the order parameter is small, the momentum fluctuations $n_{ij}$ become isotropic. However, this argument holds formally only for equilibrium, while for the non-equilibrium protocol one should evaluate the fluctuations along the whole trajectory.

To  characterize the dynamics of the fluctuations during this time-dependent process, we represent the fluctuation matrix $n_{ij}(t)$ in a polar plot, projecting it along the direction $e^R(\varphi')=(\sin(\varphi'),-\cos(\varphi'))$ in orbital pseudospin space 
\be \label{fluc_ang_distrib}
n \left(t, \varphi' \right) = e_i^{R}(\varphi') e_j^{R}(\varphi') n_{ij} \left( t \right) \;.
\ee
The function $n (t, \varphi' )$ is shown for $T=0.125$ as polar plot around several points on the trajectory (grey shaded areas). The plot shows a characteristic double lobe structure, indicating a minimum of the fluctuations in the longitudinal direction and a maximum in the transverse direction. In agreement with the entropic origin of the free energy minima, the strongest confinement of the fluctuations is observed around the thermodynamically most unstable points (middle of the trajectory in \figu{fig_ang_trajec_diff_temp}(a)). 

\begin{figure}
\centering \includegraphics[width=0.5\textwidth]{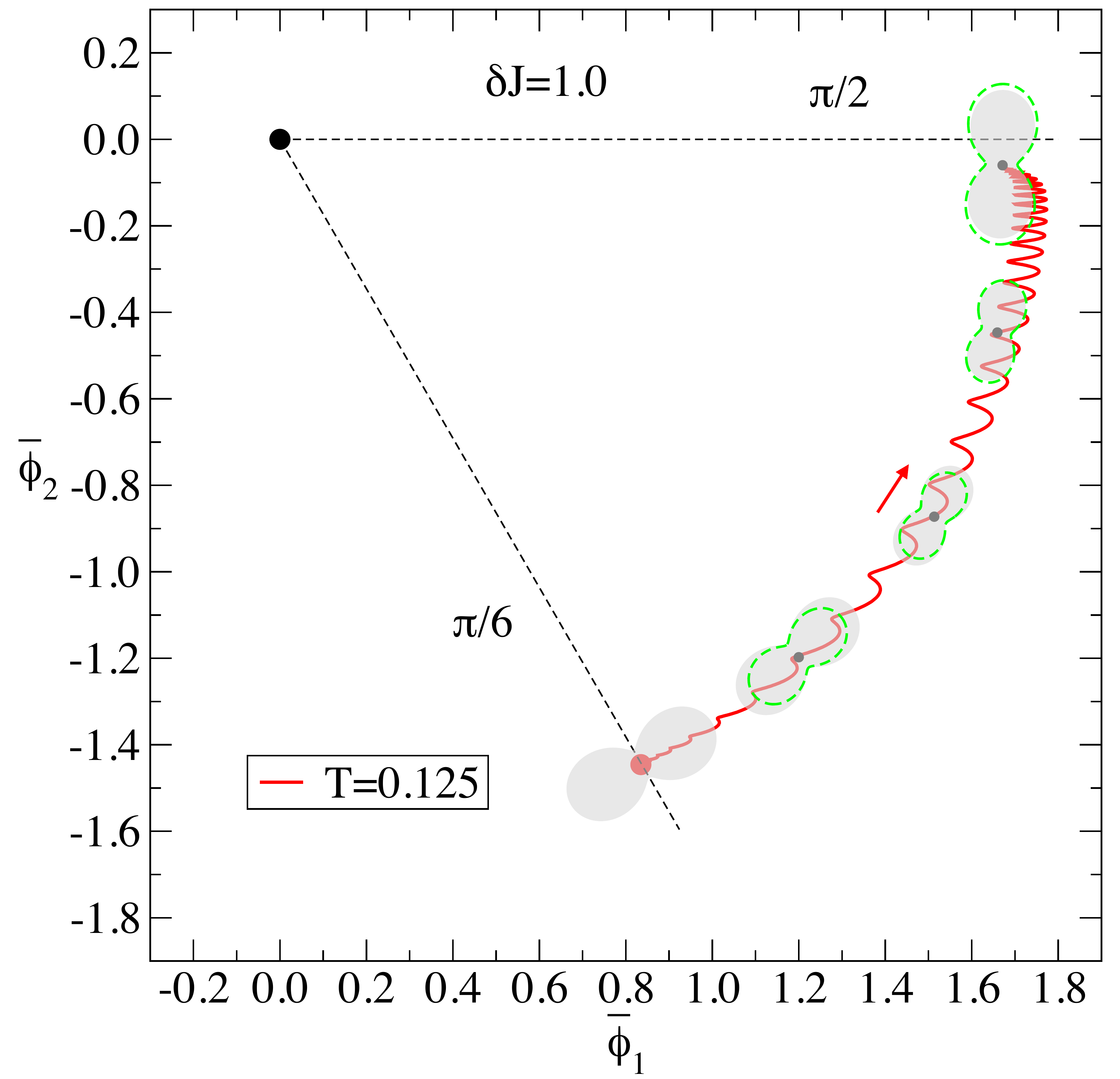}
\caption{ Trajectory at the initial temperature $T=0.125$ for the driving of the effective $J_{x}$ for pulse parameters $\delta J=1.0$ and $t_p \sim 13$. As described in Fig.~\ref{fig_ang_trajec_diff_temp}(a), the grey shaded areas and the green dashed lines compare the angular distribution of the fluctuations in nonequilibrium and the corresponding equilibrium distribution  at different points of the trajectory.
}
\label{fig_traj_strong_dJ_1}
\end{figure}

By comparison one can see that the nonequilibrium angular distribution of the fluctuations at different stages of the dynamics (shaded grey areas in \figu{fig_ang_trajec_diff_temp}(a)) and the corresponding equilibrium distribution for fixed values of the angle and the radius of the order parameter (dashed green lines in  \figu{fig_ang_trajec_diff_temp}(a)) are almost identical. This shows that the switching process is quasi-adiabatic, as far as the integrated fluctuations are concerned. The quasi-adiabatic switching can be contrasted with a process where the external electric field is substantially larger. The trajectory for a switching with $\delta J = 1.0$  (corresponding to a $20$ percent change of the exchange couplings) is shown in \figu{fig_traj_strong_dJ_1}. The correspondingly faster switching indeed leads to a mismatch between the time-dependent fluctuations and the equilibrated fluctuations at each angle $\varphi$. In general, the time-dependent fluctuations lag behind the corresponding equilibrium fluctuations.

\begin{figure*}
\centering \includegraphics[width=1\textwidth]{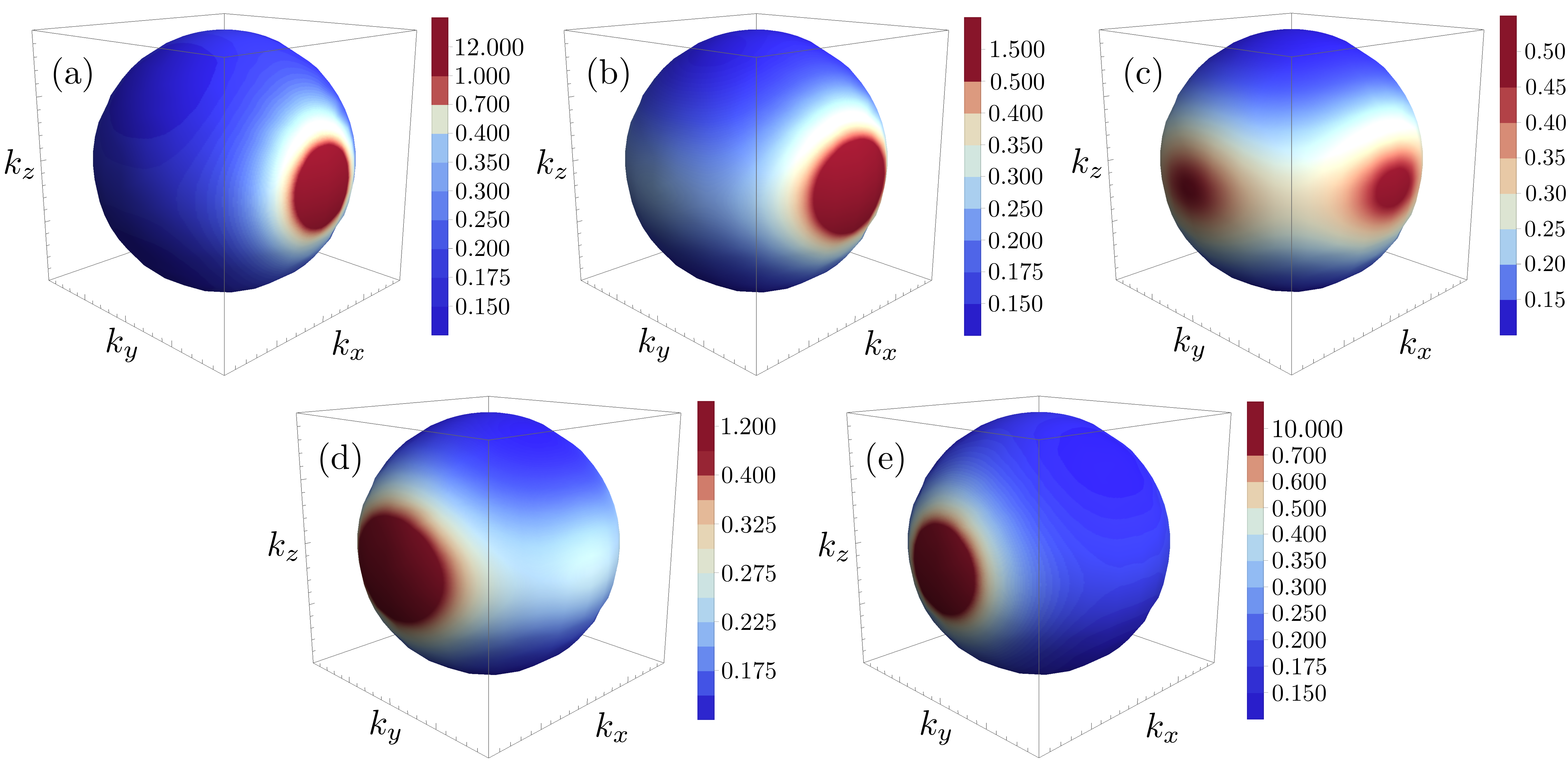}
\caption{ Spherical representation of the integral of the trace of the correlation matrix \equ{int_corr_funct} at several points of the trajectory shown in \figu{fig_ang_trajec_diff_temp}a for $T=0.125$ and $\delta J = 0.1$. The five plots correspond to the points for which the momentum integrated fluctuations are shown  in \figu{fig_ang_trajec_diff_temp}a, at (a) $\varphi =\pi/6$, (b) $\varphi = \pi/4$, (c) $\varphi = \pi/3$, (d) $\varphi = 5 \pi/12$ and (e) $\varphi \sim \pi/2$.}
\label{fig_RIXS_image}
\end{figure*}

As already the momentum averaged fluctuations show a non-trivial dynamics, it would be interesting to map out the momentum-resolved fluctuations. These are in principle accessible using scattering techniques (Sec.~\ref{secIV}), and their measurement would provide the most detailed characterization of the free energy landscape along the switching path.
In our model, the momentum-resolved fluctuations are given by the momentum distribution of the correlation matrix $D ( \mathbf{k}, t )$, whose entries are orbitally resolved. To simplify the representation of this quantity, we analyze its trace $D_{\text{Tr}}  ( \mathbf{k}, t ) = \sum_{i} D_{ii} ( \mathbf{k}, t )$. Moreover, since we are mainly interested in the directionality of the correlation function in $k$ space, we show its integral over the modulus of the wave vector,
\be \label{int_corr_funct}
D_{\text{Tr}} \left( \zeta, \chi, t \right) = \int_{0}^{\Lambda} dk \ k^{2} D_{\text{Tr}} \left( \mathbf{k}, t \right) \;,
\ee
having defined the vector $\mathbf{k} = \left( k_{x}, k_{y}, k_{z} \right)$ in spherical coordinates $k_{x} = k \sin \left( \zeta \right) \cos \left( \chi \right)$, $k_{y} = k \sin \left( \zeta \right) \sin \left( \chi \right)$ and $k_{z} = k \cos \left( \zeta \right)$. At the initial point of the dynamics, \equ{int_corr_funct} shows peaks structure along the $\pm k_x$, see \figu{fig_RIXS_image}(a). During the drag of the system from one minimum to the other, the k-resolved fluctuations \equ{int_corr_funct} continuously transfer  weight from the original maxima along $\pm k_x$ direction to the $\pm k_y$ direction, where the maxima of the final state are observed, see \figu{fig_RIXS_image}(e). At intermediate steps, the distribution of the fluctuations lies between these two limits, showing four well resolved maxima in the k distribution when the system is at the maximum of the free energy landscape, see \figu{fig_RIXS_image}(c). In \secu{secIV}, we describe a possible experimental probe to measure the $k$-resolved dynamics of the fluctuations.


\subsection{Heat-assisted switching} \label{secIIIquenchswitch}

\begin{figure}
\centering \includegraphics[width=0.5\textwidth]{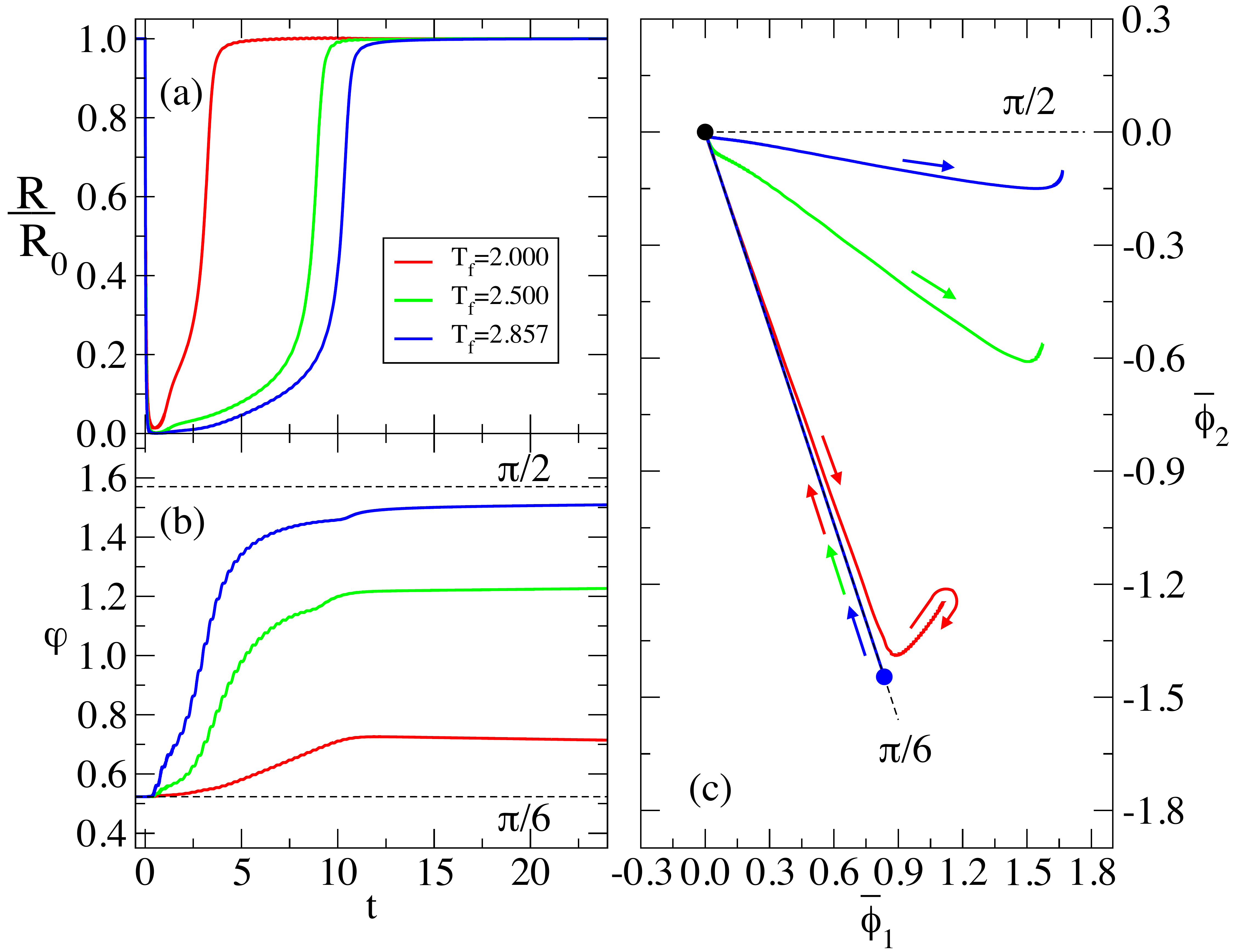}
\caption{ (a) Time-dependent change of the amplitude of the order parameter (normalized by the equilibrium value $R_{0}$) during a simultaneous  modulation of $J_{x}$ [pulse parameters $\delta J= 0.1$ and $t_p \sim 13$] and a temperature quench, for several maximum temperatures $T_{\text{f}}$.  (b) Phase of the order parameter $\varphi$ during the same time-dependent protocols. (c) Corresponding trajectories in orbital space. Arrows in panel (c) show the direction in which the state of the system moves along the trajectory.}
\label{fig_ang_rad_vs_time_comp_traj}
\end{figure}

In this section, we analyze the simultaneous action of the two time-dependent perturbations considered in \secu{secIIIquench} and \secu{secIIIswitch}. Because we have seen that in equilibrium a sufficiently high initial temperature simplifies the switching protocol, one can ask whether the switching can be facilitated  by quenching the temperature in addition to the modification of the exchange couplings. Our data confirm we can answer this question affirmatively, as we are going to show.

In \figu{fig_ang_rad_vs_time_comp_traj}, we analyze the response of the system to temperature quenches of increasing strength when we simultaneously apply a time-dependent $J_{x}$ with fixed parameters $\delta J = 0.1$ and $t_{p} \sim 13$. As apparent from \figu{fig_ang_n_cart_vs_time_comp}(a), such a short pulse without the temperature quench would not be able to induce the switch to the new orbital state. We generally notice that the presence of the time-dependent $J_{x}$ makes the high-symmetry state at $R \sim 0$ less stable, compare the time scale for the recovery in \figu{fig_ang_rad_vs_time_comp_traj}(a) and \figu{fig_temp_quench_rT_comp}. For small quenches in temperature, the dynamics of the phase of the order parameter is not substantially perturbed to the case without the quench, see \figu{fig_ang_rad_vs_time_comp_traj}(b) for $T_{\text{f}} = 2.000$. In this case, $\varphi$ first increases due to the action of $J_{x} (t)$, but it slowly recovers the original value $\varphi = \pi/6$ once the pulse is over. If the quench is sufficiently strong, the stability of the $R \sim 0$ state lasts enough for the time dependent $J_{x}$ to make the switch, as observed in \figu{fig_ang_rad_vs_time_comp_traj}(b) for $T_{\text{f}} = 2.857$. Trajectories for different quenches are shown in \figu{fig_ang_rad_vs_time_comp_traj}(c).

This result is akin to the mechanism observed in heat-assisted recording \cite{4694026}, where the heating of magnetic memory devices reduces the size of the magnetic field which is needed to switch the magnetization state of the system against some anisotropy. However, there are some important differences (apart from the fact that orbital order responds to different forces). Firstly, the order-by-disorder mechanism implies that there is a threshold for the temperature quench below which it would no longer facilitate the switching, because for low temperatures the restoring forces would increase with increasing fluctuations instead of decreasing. Moreover, we emphasize that the transient stabilization of the high-symmetry state ($R \sim 0$) by the fluctuations also simplifies the switching because this prolongs the time in which the order parameter can be manipulated with relatively weak fields.


\section{Measurement of the fluctuations} \label{secIV}

In this section, we briefly discuss the measurement of the $k$-resolved fluctuation spectrum based on X-ray scattering. RIXS can give insights into orbital correlations in the ground state of correlated materials \cite{PhysRevLett.101.106406,PhysRevLett.109.117401,RevModPhys.83.705}. Time-resolved RIXS, based on free-electron laser sources, is therefore a potentially very versatile probe for complex solids out of equilibrium \cite{Mitrano2020}. In RIXS, an incoming X-ray photon with energy $\omega$ and momentum $\mathbf{k}$ excites a core hole that subsequently decays through the emission of an outgoing photon with energy $\omega'$ and momentum $\mathbf{k}'$. The excess energy $\Omega = \omega - \omega'$ and momentum $\mathbf{q} = \mathbf{k} - \mathbf{k}'$ are transferred to the system, which allows to explore the energy momentum dispersion of the low energy excitations. The polarization $\boldsymbol{\epsilon}$ and $\boldsymbol{\epsilon}'$ of the ingoing and outgoing photons can be used to gain orbital selectivity. In general, under the assumption of a localized core hole with a short lifetime $1/\Gamma$, the scattering can be expressed in terms of a structure factor for transition operators $A$ and $B$ acting in the low-energy manifold of the solid (valence excitations without core-hole),
\be \label{struct_fact}
S_{A,B} \left( \mathbf{q}, \Omega \right) 
\! =\! \sum_{f} \langle g \vert A_{\mathbf{q}}^{\dagger} \vert f \rangle \langle f \vert B_{\mathbf{q}} \vert g \rangle \delta( \Omega + E_{g} - E_{f} ).
\ee
Here $\vert g \rangle$, with energy $E_{g}$, is the initial ground state of the system, and $|f\rangle$ an excited state in the low-energy valence manifold. (The ground state average is replaced by a statistical average for finite temperature.) Within this approximation, the RIXS intensity for scattering between different orbital states in the $e_g$ manifold has been expressed in terms of the orbital structure factor \cite{marra2016theoretical},
\begin{align}
&I_{\text{RIXS}}  \propto \left\vert W^{A}_{\boldsymbol{\epsilon}, \boldsymbol{\epsilon}'} \right\vert^{2} S_{\tau^{3},\tau^{3}} \left( \mathbf{q}, \Omega \right) +  \left\vert W^{O}_{\boldsymbol{\epsilon}, \boldsymbol{\epsilon}'} \right\vert^{2} S_{\tau^{1},\tau^{1}} \left( \mathbf{q}, \Omega \right) 
\nonumber\\
 \label{scatt_cross_sec_q_structure}
& + \left[ ( W^{A}_{\boldsymbol{\epsilon}, \boldsymbol{\epsilon}'} )^{*} W^{O}_{\boldsymbol{\epsilon}, \boldsymbol{\epsilon}'} + W^{A}_{\boldsymbol{\epsilon}, \boldsymbol{\epsilon}'} ( W^{O}_{\boldsymbol{\epsilon}, \boldsymbol{\epsilon}'} )^{*} \right] S_{\tau^{1},\tau^{3}} \left( \mathbf{q}, \Omega \right),
\end{align}
with the polarization-dependent matrix elements,
\be \label{polarizations_pre}
\begin{split}
& W^{A}_{\boldsymbol{\epsilon}, \boldsymbol{\epsilon}'} = \frac{2}{i \Gamma} \left( \frac{1}{3} \boldsymbol{\epsilon} \cdot \boldsymbol{\epsilon}' - \epsilon_{x} \epsilon_{x} '^{*} \right) \;, \\
& W^{O}_{\boldsymbol{\epsilon}, \boldsymbol{\epsilon}'} = - \frac{1}{\sqrt{3} i \Gamma} \left( \epsilon_{y} \epsilon_{y} '^{*} - \epsilon_{z} \epsilon_{z} '^{*} \right) \;.
\end{split}
\ee
By properly choosing the incoming and the outgoing polarization of the photons, one can therefore select specific components of the total RIXS intensity. For instance, by selecting $\boldsymbol{\epsilon}$ and $\boldsymbol{\epsilon}'$ such that $\frac{1}{3} \boldsymbol{\epsilon} \cdot \boldsymbol{\epsilon}' = \epsilon_{x} \epsilon_{x} '^{*}$ and $\epsilon_{y} \epsilon_{y} '^{*} \neq \epsilon_{z} \epsilon_{z} '^{*}$, we obtain $I_{\text{RIXS}} \propto S_{\tau^{1},\tau^{1}} \left( \mathbf{q}, \Omega \right)$. Similarly, we can obtain the other diagonal component $S_{\tau^{3},\tau^{3}} \left( \mathbf{q}, \Omega \right)$ and the off-diagonal component  $S_{\tau^{1},\tau^{3}} \left( \mathbf{q}, \omega \right)$.

The static correlation functions discussed in \secu{secIIIswitch} are then obtained by an energy integration in the energy window relevant for the orbital excitations,
\be \label{corr_struct_fact}
D_{a,b} \left( \mathbf{k} \right) \propto \int d \Omega \ S_{A,B} \left( \mathbf{k}, \Omega \right) \;,
\ee
with $a,b = 1,2$, $A,B = \tau^{3}, \tau^{1}$ and the correspondences $1 \leftrightarrow \tau^{3}$, $2 \leftrightarrow \tau^{1}$. Thus, through diffuse X-ray scattering, one can probe the momentum distribution of the correlation functions in a highly anisotropic system such as a compass model. Because the signal is energy integrated, it should generalize directly to time-resolved probes, with the same selection rules for the different components of the orbital fluctuations.


\section{Conclusions} \label{secV}

In this work, we designed a time-dependent Ginzburg-Landau formalism suitable to study the non-equilibrium dynamics of the $120^{\circ}$ compass model, as a paradigmatic representative for a system in which a multi-minima free energy landscape for orbital order arises from the order by disorder scenario. We find that the entropic forces that determine the free energy minima at equilibrium play an important role also under nonequilibrium conditions: There is a relatively long-time stabilization of the otherwise unstable unbroken symmetry state once non-thermal fluctuations of the order parameters have been activated. Moreover, we find it is generally possible to switch from one minimum to another considering a time dependent exchange coupling that emulates the action of a laser-pulse. If, together with this time dependent exchange interaction, we consider a temperature quench, the orbital switching is speeded up by one order of magnitude even for small applied external fields. This mechanism can be viewed as a generalization of the heat-assisted switching in magnetic memory devices to the orbitally-ordered states. 

The present analysis is certainly highly idealized. For generic materials, one cannot neglect the dynamics of the spin, and further parameters beyond the effective temperature $T$ may be needed to describe the non-equilibrium state of the electronic degrees of freedom, in particular for insulators \cite{Beaud2014}. Moreover, microscopic theory (based on spin models, or the full electronic model) would be needed to get an estimate of phenomenological constants, such as the damping, and also to understand the  ultra-fast regime in which even the ``fast''  degrees of freedom are not thermalized \cite{PhysRevB.103.035116}. Nevertheless, general aspects discussed in the present study should carry over to more realistic systems: 

(1) In particular, the present work shows that for systems which follow the order by disorder mechanism in equilibrium, forces arising from fluctuations should also be taken into account in the interpretation of their dynamics. Corrections to the free energy potential from non-thermal fluctuations become increasingly more important for faster processes, as in the investigation of the non-adiabatic switching in Fig.~\ref{fig_traj_strong_dJ_1}.
(2) While control parameters such as the electron temperature $T$ and order parameter fluctuations are rigidly linked in equilibrium, they independently control the dynamics out of equilibrium. In the present model the manifestation of this is relatively simple (non-thermal fluctuations stabilize the symmetry unbroken state even when $T$ is below the critical temperature).

This fact motivates a search for situations in which other, nontrivial hidden states can be transiently stabilized by nonthermal fluctuations. Such phenomena should be investigated, ideally by using scattering techniques to analyze the time-dependent orbital fluctuations, in transition metal compounds such as KCuF$_{3}$. Estimates for a two-band Hubbard model suggest that for Mott-based orbitally ordered systems electric field-induced changes of the exchange interaction up to few percent should be possible \cite{eckstein2017designing}, potentially allowing to drive coherent dynamics of the order parameter.

\section*{Acknowledgment}
We gratefully acknowledge the work done by Aaron M\"uller at the early stages of the project.
We were supported by the ERC Starting Grant No. 716648.


\appendix

\section{Equilibrium solution of the problem}  \label{app_a}
Under equilibrium conditions, the solution of \equ{dyn_eq_corr} for the correlation functions reads:
\be
D_{i j} \left( \mathbf{k}, t \right) = \left( -1 \right)^{i+j} \frac{T M_{\bar{i} \bar{j}} \left( \mathbf{k}, t \right)}{\text{det} \left[ M \left( \mathbf{k}, t \right) \right]} \;,
\ee
where the $M \left( \mathbf{k}, t \right)$ matrix has elements as defined in \equ{m_matrix}, so:
\be \label{denominator}
\text{det} \left[ M \left( \mathbf{k}, t \right) \right] = M_{11} \left( \mathbf{k}, t \right) M_{22} \left( \mathbf{k}, t \right) - M_{12}^{2} \left( \mathbf{k}, t \right) \;,
\ee
and $\bar{i} = 3 - i$. Here, we analyze the condition where $J_{x}=J_{y}=J_{z} \equiv J$, leading to $A_{11}^{0} = A_{22}^{0} \equiv A^{0} = \frac{3}{2} r \left( T \right) J$ and $A_{12}^{0} = 0$. In passing, we notice that, along specific directions in $k$ space, the correlation function does not decay to zero when $\vert \mathbf{k} \vert \rightarrow \infty$, making the fluctuations \equ{exc_num} cutoff dependent. However, the physics we describe is qualitatively independent of the choice of $\Lambda$. We distinguish two cases in our analysis, i.e. the zero-temperature condition $T=0$ and the limit of high temperatures. 

At zero temperature, all fluctuations vanish, so that the effective quadratic couplings \equ{r_param_eff} read $A^{\text{eff}}_{11} = A^{\text{eff}}_{22} = A^{\text{eff}} = A^{0} + 4 u \left( \bar{\phi}_{1}^{2} + \bar{\phi}_{2}^{2} \right)$ and $A^{\text{eff}}_{12} = 0$. The stationarity condition for \equ{dyn_eq} implies that $A^{\text{eff}} = 0$, thus $\bar{\phi}_{1}^{2} + \bar{\phi}_{2}^{2} = - \frac{A^{0}}{4u}$, reflecting the $\text{O} \left( 2 \right)$ symmetry of the zero-temperature problem. At high temperatures $T > T_{c}$, the order parameter is equal to zero, so $\bar{\phi}_{1} = \bar{\phi}_{2} = 0$. In this case, the momentum-integrated fluctuations become isotropic, $n_{12} = 0$ and $n_{11} = n_{22} = n$, so that $A^{\text{eff}}_{11} = A^{\text{eff}}_{22} = A^{0} + 16 u n$.



%

\end{document}